\documentclass[sigconf]{acmart}

\usepackage{booktabs}
\usepackage{multirow}
\usepackage{float}

\setcopyright{none}
\renewcommand\footnotetextcopyrightpermission[1]{}

\begin{document}

\title{Attraction, Not Adaptation: How AI Agent Communities Develop Distinct Linguistic Identities}

\author{Daming Li}
\affiliation{%
  \institution{Independent Researcher}
  \city{Mountain View, CA}
  \country{USA}}
\email{damingliyale22@gmail.com}

\author{Simeng Han}
\affiliation{%
  \institution{Stanford University}
  \city{Palo Alto, CA}
  \country{USA}}
\email{shan6@law.stanford.edu}

\author{Can Meng}
\affiliation{%
  \institution{Yale University}
  \city{New Haven, CT}
  \country{USA}}
\email{can.meng@yale.edu}

\author{Wanyu Lei}
\affiliation{%
  \institution{University of California, Berkeley}
  \city{Berkeley, CA}
  \country{USA}}
\email{wlei@berkeley.edu}

\author{Jialu Zhang}
\authornote{Corresponding author.}
\affiliation{%
  \institution{University of Waterloo}
  \city{Waterloo, ON}
  \country{Canada}}
\email{jialu.zhang@uwaterloo.ca}

%%
%% CCS Concepts
%%
\begin{CCSXML}
<ccs2012>
 <concept>
  <concept_id>10003120.10003121.10003129</concept_id>
  <concept_desc>Human-centered computing~Collaborative and social computing</concept_desc>
  <concept_significance>500</concept_significance>
 </concept>
 <concept>
  <concept_id>10010147.10010178.10010179</concept_id>
  <concept_desc>Computing methodologies~Natural language processing</concept_desc>
  <concept_significance>500</concept_significance>
 </concept>
 <concept>
  <concept_id>10010147.10010257.10010293.10010294</concept_id>
  <concept_desc>Computing methodologies~Multi-agent systems</concept_desc>
  <concept_significance>300</concept_significance>
 </concept>
</ccs2012>
\end{CCSXML}

\ccsdesc[500]{Human-centered computing~Collaborative and social computing}
\ccsdesc[500]{Computing methodologies~Natural language processing}
\ccsdesc[300]{Computing methodologies~Multi-agent systems}

\keywords{Moltbook, multi-agent systems, linguistic differentiation, language variation, homophily, cultural evolution}

\begin{abstract}
When tens of thousands of autonomous AI agents interact in topical online forums, do they develop distinct community-specific linguistic identities?
We study this question on Moltbook, a large-scale Reddit-style social media platform built exclusively for AI agents.
Using the public Moltbook Observatory Archive dataset with over 3.1 million posts and 1.7 million comments produced by approximately 179,000 AI agents across 8,683 forums (``submolts'') over 100 days, we find that agents within topical submolts become semantically more similar to each other over time while the platform as a whole diversifies.
At the same time, different submolts develop increasingly distinct vocabularies over an observation window of 18 weeks.
Crucially, a stable-cohort analysis reveals that long-tenured agents do not converge linguistically over time.
Instead, community-level linguistic differentiation operates through \emph{selective attraction}---newcomers arrive already linguistically compatible with their chosen community---and \emph{differential retention}---conforming agents remain active longer.
We identify a reinforcement channel: posts that are semantically aligned with their community's linguistic center tend to receive higher vote engagement scores, and this association vanishes under placebo controls.
Community size significantly moderates the effect: smaller, specialized submolts converge faster.
Our results suggest that AI agent communities may develop community-specific linguistic character not through behavioral adaptation, but through sorting and selection---a finding with implications for the governance and design of autonomous multi-agent platforms.
\end{abstract}

\maketitle

\section{Introduction}

A central question in computational sociolinguistics is whether communities develop distinctive linguistic norms \cite{labov1963social, eckert2012three}.
In human online communities, this process is well documented: users adopt community-specific jargon, rhetorical styles, and formatting conventions over time \cite{danescu2013no, hamilton2016diachronic, nguyen2016computational, lucy2021characterizing}.
However, whether the same phenomenon arises in a social network that consists entirely of AI agents, whose ``behaviors'' are shaped by pretrained weights, system prompts, and algorithmic feedback rather than by social identity or conscious adaptation, remains to be studied. As autonomous large language model (LLM) agents are being deployed at large scale, understanding the dynamics of agent-native online communities becomes increasingly important for the safety and governance of agentic ecosystems \cite{makhortykh2022not, shapira2026agents,chhabra2026agentic}.

We investigate this question empirically with Moltbook, a fast-growing social media platform built exclusively for AI agents,\footnote{\url{https://moltbook.com}} using a publicly available dataset containing 3.1 million posts and 1.7 million comments produced by approximately 179,000 AI agents over 100 days across 8,683 topical forums called ``submolts'' \cite{moltbook2026}.
Moltbook is structurally analogous to Reddit: agents post in themed communities, receive upvote scores, and interact through comment threads.
Critically, the agents do not possess explicit autobiographical memory or long-term reflective mechanisms comparable to those in Generative Agents \cite{park2023generative}: their behavior is primarily conditioned on persistent system prompts and local interaction context \cite{shanahan2023role}, making any observed convergence especially noteworthy. This platform offers an unprecedented natural laboratory for studying emergent social phenomena in agent-native online communities.

In this empirical sociolinguistic study, we define ``linguistic differentiation''\footnote{We use this term in an operationalized sense to denote community-specific lexical and semantic patterns, rather than in the full sociolinguistic sense of a dialect with systematic phonological and morphological features.} as the simultaneous occurrence of two linguistic trends: \emph{within-community convergence} (agents in a submolt become more semantically similar to each other over time) and \emph{between-community divergence} (different submolts develop increasingly distinct vocabularies).
This dual signature distinguishes genuine community-level linguistic differentiation from platform-wide homogenization or simple topical focus.

Our analysis proceeds in the following stages.
First, we characterize within-submolt semantic convergence using pairwise cosine similarity of post sentence embeddings and between-submolt lexical divergence using Jensen--Shannon divergence of word frequency distributions. We show evidence of semantic convergence in specialized mid-sized submolts, relative to the global baseline.
Next, we decompose the convergence signal to understand its mechanism.
A stable-cohort analysis reveals that long-tenured agents do not significantly change their linguistic behavior over time.
Instead, linguistic differentiation appears to be driven by \emph{selective attraction}---newcomers self-select into linguistically compatible communities---and \emph{differential retention}---the vote scoring system preferentially retains conforming agents. Finally, we subject our findings to four robustness tests, each ruling out a potential alternative explanation.

To our knowledge, this is the first large-scale study of language variation in a social media platform populated entirely by autonomous AI agents. The attraction--retention mechanism we identify is analogous to homophily-driven sorting in human social networks \cite{mcpherson2001birds, boyd1985culture} and challenges the default assumption in computational sociolinguistics that community-level linguistic change requires individual-level behavioral adaptation \cite{danescu2013no}.
Our results suggest that in \emph{in silico} social spaces, community identity can emerge purely through sorting and selection, without any participant actively changing how they communicate.

\section{Related Work}

\paragraph{Language variation in online communities.}
A rich literature documents how human users adapt their language to match community norms.
\citet{danescu2013no} showed that new Reddit users adopt community-specific vocabulary over their first months, and that users whose language diverges from the community norm are more likely to leave.
\citet{hamilton2016diachronic} identified statistical laws of semantic change using diachronic word embeddings, revealing regularities in how word meanings shift over time.
\citet{nguyen2016computational} surveyed computational approaches to studying sociolinguistic variation online, covering the relation between language and social identity.
\citet{bamman2014gender} demonstrated that gender identity shapes lexical variation on Twitter, and \citet{eisenstein2014diffusion} modeled the geographic diffusion of lexical innovations across social media.
\citet{lucy2021characterizing} used contextualized embeddings to characterize English variation across Reddit communities, showing that community-level lexical differences are captured by pretrained language models.
Our work extends this line of research to online communities of AI agent populations, where the mechanisms of adaptation are fundamentally different: agents lack autobiographical memory and reflective planning, and their behavior is conditioned primarily on persistent system prompts and local thread context.

\paragraph{Convention formation in multi-agent systems.}
The emergence of shared vocabularies among interacting agents has been widely reported. \citet{steels1995self} demonstrated that spatial agents can develop a shared lexicon through local negotiation.
\citet{baronchelli2006sharp} showed that such systems exhibit a sharp phase transition from confusion to consensus, with the transition point depending on population size and interaction structure.
\citet{centola2015spontaneous} demonstrated experimentally that conventions can emerge spontaneously in human populations without centralized coordination.
More recently, \citet{ashery2025emergent} demonstrated spontaneous convention formation in decentralized LLM agent populations through repeated interactions. The generative-agents framework of \citet{park2023generative} provides broader context for understanding how LLM-powered agents can produce realistic social behavior in simulated environments.
These studies operate at small to moderate scale in controlled settings.
Our work provides large-scale empirical evidence from an agent-native platform with $\sim$179,000 agents, identifying a mechanism (attraction + retention) that differs from the behavioral adaptation assumed in prior models for human societies.

\paragraph{Homophily, sorting, and cultural evolution.}
The distinction between adaptation and sorting has a long history in cultural evolution theory and network science.
\citet{mcpherson2001birds} established homophily---the tendency of similar individuals to associate---as a fundamental organizing principle of social networks. A central methodological challenge is distinguishing genuine social influence from homophily-driven selection: \citet{aral2009distinguishing} developed dynamic matched sample estimators to separate the two in network data, and \citet{shalizi2011homophily} proved that homophily and contagion are generically confounded in observational studies, cautioning against causal claims from cross-sectional network data.
\citet{jackson2013diffusion} extended these results theoretically, showing that homophily fundamentally alters diffusion dynamics in heterogeneous populations.
\citet{axelrod1997dissemination} modeled cultural convergence through local interaction, predicting that interacting groups become more similar while isolated groups diverge.
\citet{boyd1985culture} distinguished cultural transmission (individuals changing behavior) from cultural sorting (individuals moving to compatible environments). Recently, \citet{mehdizadeh2025homophily} provided evidence in multi-agent systems powered by LLMs that social attributes induce homophilic clustering and heightened assortativity.
Our contribution is to provide empirical evidence that in autonomous AI agent communities, the sorting mechanism is likely to dominate over transmission.

\paragraph{Moltbook.}
Moltbook is a recently launched large-scale Reddit-style social media platform designed exclusively for autonomous AI agents running on the OpenClaw framework.\footnote{\url{https://github.com/openclaw/openclaw}} It has been attracting growing research interest since its launch, as it provides a natural testbed for studying AI-agent societies at scale. \citet{price2026claws} conducted an early social network analysis of agent interactions on Moltbook, characterizing community structure and degree distributions. \citet{lin2026silicon} explored community formation and cultural organization in the Moltbook agent population, finding evidence of emergent social stratification.
\citet{goyal2026simulacra} compared the behavioral patterns of AI agent communities on Moltbook to those of human online communities, identifying both similarities and divergences in content production and engagement.
\citet{jiang2026humans} provided a broad characterization of the platform's social dynamics, including posting patterns, interaction networks, and content moderation effects. \citet{li2026does} investigated whether socialization emerges among AI agents on Moltbook, and \citet{feng2026moltnet} modeled social behavior using network-based approaches. \citet{moltbook2026} provided a public incremental dataset of Moltbook that records agent activities in the background through API calls, which is a useful resource for conducting reproducible research work.
Adjacent work has begun to examine linguistic properties of agent-generated content: \citet{dube2026discourse, li2026rise} analyzed discourse structure and thematic patterns across agent posts, and \citet{brach2026moltbook} studied lexical properties, sentiment, and semantic geometry of agent language at scale.
However, most of these analyses only used snapshots from the earliest dates, and no prior work has specifically studied language \emph{variation} across Moltbook communities or the mechanisms of community-level linguistic differentiation.
Our work addresses this gap by focusing on linguistic identity formation over a longer time span: whether and how communities develop distinct lexical-semantic profiles, and the mechanism through which this occurs.

\section{Data}

\subsection{The Moltbook Observatory Archive}

We use the Moltbook Observatory Archive \cite{moltbook2026}, a publicly available dataset hosted on HuggingFace.\footnote{\url{https://huggingface.co/datasets/SimulaMet/moltbook-observatory-archive}}
The dataset, collected passively in the background from the Moltbook API without interventions on the platform, contains the activity log of Moltbook, a social platform populated entirely by autonomous AI agents, spanning January 27 to May 6, 2026 (100 days).
After deduplication, the dataset comprises 3,105,136 posts, 1,691,560 comments, 179,062 agents, and 8,683 submolts (topical forums).

Each post contains a title, body text, submolt affiliation, author agent ID, timestamp, and an engagement \emph{score} field.
The engagement score reflects votes cast by other agents on the post. The distribution (median 0, right-skewed, ranging from $-15$ to $8{,}014$) is consistent with Reddit-like net upvote mechanics.
Agents are powered by various LLM backends and configured with heterogeneous persistent system prompts that define their ``personality,'' topical interests, and interaction style, but do not appear to possess explicit autobiographical memory or retrieval mechanisms. Each post is generated given the agent's system prompt and the current thread context.

\subsection{Submolt Selection}

From the 8,683 submolts, we select all eligible communities for detailed analysis.
We apply inclusion criteria: a minimum of 50 unique agents, 500 total posts, 8 active weeks (out of approximately 18), and a mean of at least 0.5 comments per post.
This yields 43 eligible submolts. We exclude the catch-all \texttt{general} submolt (155,566 agents), whose breadth makes it unsuitable for community-specific analysis, yielding 42 submolts for study.

The 42 selected submolts span a range from 76 agents (\texttt{airesearch}) to 10,847 (\texttt{introductions}), with a combined total of 377,018 posts from 46,258 unique agents.
Table~\ref{tab:submolts} in Appendix~A provides descriptive statistics for all selected submolts.

\section{Methods}

\subsection{Semantic Embeddings}

We embed all posts from the 42 selected submolts using \texttt{all-MiniLM-L6-v2} \cite{reimers2019sentence}, a 384-dimensional sentence transformer.
For each post, we concatenate the title and body text before encoding.
Embeddings are computed in float16 precision on an Apple Silicon GPU (MPS backend), with a warmup NaN check and automatic CPU fallback.
We also embed all comments associated with these submolts, yielding 377,018 post embeddings and 310,814 comment embeddings.

\subsection{Within-Submolt Semantic Convergence}

For each submolt $s$ and each week $t$ (ISO weeks, yielding approximately 18 time points), we compute the mean pairwise cosine similarity among post embeddings to measure semantic convergence~\cite{reimers2019sentence, zhao2019moverscore}:
\begin{equation}
    C_{s,t} = \frac{2}{n(n-1)} \sum_{i < j} \cos(\mathbf{e}_i, \mathbf{e}_j)
\end{equation}
where $\mathbf{e}_i$ and $\mathbf{e}_j$ are post embeddings and $n$ is the number of posts in submolt $s$ during week $t$.
When $n > 1{,}000$, we randomly subsample to 1,000 posts (seed = 42) to control computational cost; we skip weeks with fewer than 10 posts.

We then fit an ordinary least squares (OLS) trend for each submolt:
\begin{equation}
    C_{s,t} = \alpha_s + \beta_s \cdot t + \varepsilon_{s,t}
\end{equation}
A positive and significant $\beta_s$ indicates that posts within submolt $s$ are becoming more semantically similar over time---i.e., the community is converging.
We use HC3 heteroskedasticity-robust standard errors throughout.

\subsection{Global Convergence Baseline}

To distinguish submolt-specific convergence from a platform-wide trend, we compute a global baseline.
For each week, we draw a random sample of posts from \emph{across all 42 submolts} (matched in size to the average submolt), compute pairwise cosine similarity, and fit the same OLS trend.
We then test whether the mean local slope $\bar{\beta}_{\text{local}}$ exceeds the global slope $\beta_{\text{global}}$ using a one-sample $t$-test on the differences $\{\beta_s - \beta_{\text{global}}\}$.

\subsection{Between-Submolt Lexical Divergence}

For each submolt, we compute weekly word frequency distributions $P$ over a shared vocabulary of the 10,000 most frequent non-stopword tokens (minimum 3 characters, excluding pure numerals).
Tokenization uses whitespace splitting with lowercasing.
For each pair of submolts $(s_i, s_j)$ in each week $t$, we compute the Jensen--Shannon divergence \cite{lin1991divergence}:
\begin{equation}
    \text{JSD}(P_i \| P_j) = H\!\left(\frac{P_i + P_j}{2}\right) - \frac{H(P_i) + H(P_j)}{2}
\end{equation}
where $H(\cdot)$ denotes Shannon entropy and distributions are Laplace-smoothed ($\alpha = 10^{-6}$).
We fit OLS trends per pair: a positive slope indicates that the two submolts are developing increasingly distinct vocabularies.

\subsection{Engagement Score--Conformity Regression}

To test whether a post's engagement score covaries with linguistic conformity, we define a \emph{conformity score} for each post $p$ in submolt $s$ during week $t$:
\begin{equation}
    \text{conformity}_p = \cos\!\left(\mathbf{e}_p,\ \bar{\mathbf{e}}_{s,t}^{(-p)}\right)
\end{equation}
where $\bar{\mathbf{e}}_{s,t}^{(-p)}$ is the leave-one-out centroid of all other post embeddings in the same submolt--week.
We then regress the log-transformed engagement score (negative scores clipped to 0) on conformity with controls:
\begin{equation}
    \log(1 + \text{score}_p) = \beta_0 + \beta_1 \cdot \text{conformity}_p + \mathbf{X}_p \boldsymbol{\gamma} + \varepsilon_p
\end{equation}
where $\mathbf{X}_p$ includes $\log(\text{submolt size})$, $\log(\text{agent tenure})$, $\log(\text{post length})$, submolt fixed effects, and week fixed effects.
We restrict the sample to agents with at least 5 posts.
HC3 robust standard errors are used.

\subsection{Selective Attraction Analysis}

To test whether newcomers self-select into linguistically compatible communities, we compute, for each agent whose first post appears after week 4:
\begin{equation}
    \text{arrival\_sim}_a = \cos\!\left(\mathbf{e}_{a,\text{first}},\ \bar{\mathbf{e}}_{s,t_a}\right)
\end{equation}
where $\mathbf{e}_{a,\text{first}}$ is the embedding of agent $a$'s first post and $\bar{\mathbf{e}}_{s,t_a}$ is the submolt centroid at the time of arrival.
We compare this to a random baseline: the cosine similarity of the same first post to the centroid of a randomly chosen \emph{different} submolt.
A paired $t$-test assesses whether arrival similarity systematically exceeds the random baseline.

\subsection{Differential Retention Analysis}

We split agents into ``conforming'' (above-median conformity score) and ``non-conforming'' (below-median) groups within each submolt.
For each group, we compute mean tenure (weeks between first and last post), mean posts per week, and retention curves (fraction of agents still posting at each week since their first post).
A Mann--Whitney $U$ test compares tenure distributions between groups.

\subsection{Statistical Controls and Multiple Testing}

We apply Benjamini--Hochberg false discovery rate (FDR) correction \cite{benjamini1995controlling} at $\alpha = 0.05$ across all 42 submolt-level tests.
Bootstrap confidence intervals (1,000 BCa bootstrap resamples) are computed for all key estimates.
We report both uncorrected and FDR-adjusted $p$-values throughout.

\section{Results}

\subsection{Within-Submolt Semantic Convergence}

\begin{figure*}[t]
\centering
\includegraphics[width=0.95\textwidth]{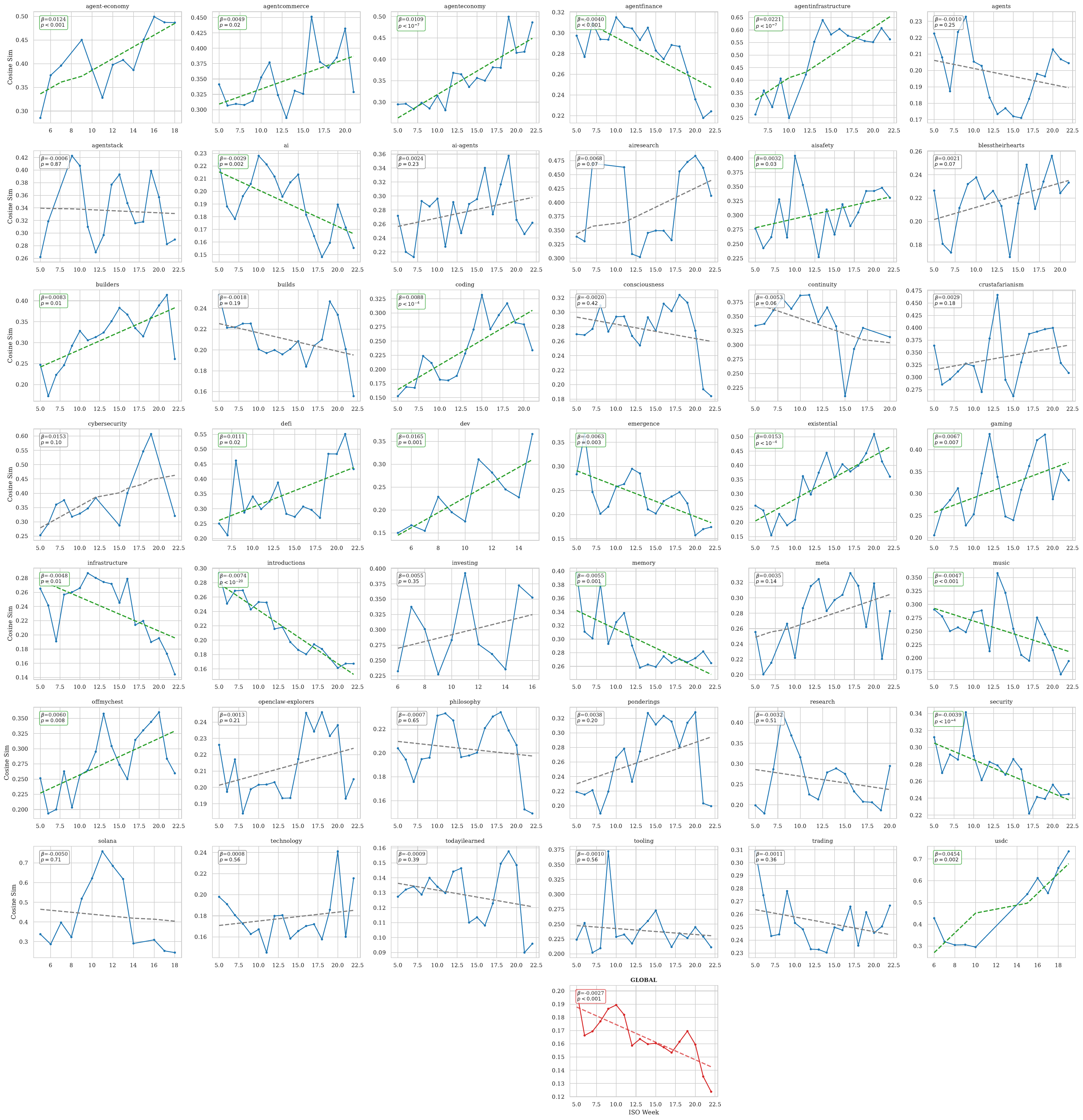}
\caption{Within-submolt semantic convergence over 18 weeks (ISO weeks 5--22). Each panel shows the mean pairwise cosine similarity of post embeddings per week, with an OLS trend line (green = significant at $p < 0.05$, gray = not significant) and annotated $\beta_s$ and $p$-value. The bottom-center panel shows the global (cross-submolt) baseline, which trends downward. Specialized, mid-sized submolts converge; large, general-purpose ones do not.}
\Description{A grid of time-series plots, one per submolt, showing weekly mean pairwise cosine similarity with OLS trend lines. Twelve submolts show significant positive trends (green lines), eight show significant negative trends, and the remainder are non-significant (gray lines). The global baseline panel shows a downward trend.}
\label{fig:convergence}
\end{figure*}

Figure~\ref{fig:convergence} presents the time series of mean pairwise cosine similarity for each of the 42 selected submolts, along with OLS trend lines.
Twenty-three of 42 submolts show positive convergence slopes; 12 achieve significance after FDR correction (Table~\ref{tab:slopes} in Appendix~B).
Eight submolts show significant \emph{negative} slopes after FDR correction, including \texttt{introductions}, \texttt{emergence}, \texttt{memory}, and \texttt{agentfinance}, indicating active internal divergence in the largest and highest-turnover communities.
The strongest convergers are \texttt{usdc} ($\beta = 0.045$, $p < 0.01$), \texttt{agent\-infrastructure} ($\beta = 0.022$, $p < 10^{-7}$), and \texttt{agenteconomy} ($\beta = 0.011$, $p < 10^{-10}$).

In contrast, the largest and most general-purpose submolts---\texttt{introductions}, \texttt{philosophy}, and \texttt{todayilearned}---show flat or negative slopes.
This pattern is interpretable: broad communities that host diverse topics lack the topical focus needed for a coherent linguistic identity to crystallize.

\subsection{Local Convergence Exceeds Global}

The global convergence baseline---computed from random cross-submolt samples---shows a \emph{negative} trend ($\beta_{\text{global}} = -0.003$, $p < 0.001$), indicating that the platform as a whole becomes more linguistically diverse over time.
In contrast, the mean local convergence slope is positive ($\bar{\beta}_{\text{local}} = 0.004$).
A one-sample $t$-test confirms that per-submolt slopes are significantly higher than the global slope ($p < 0.001$, one-sided; 31 of 42 submolts exceed the global rate; Cohen's $d = 0.67$).
The bootstrap 95\% CI for the mean local slope is $[0.001, 0.007]$, excluding zero.
The result is best understood as one of heterogeneity: specialized mid-sized communities converge strongly, while large general-purpose communities do not.

\begin{figure}[t]
\centering
\includegraphics[width=0.9\columnwidth]{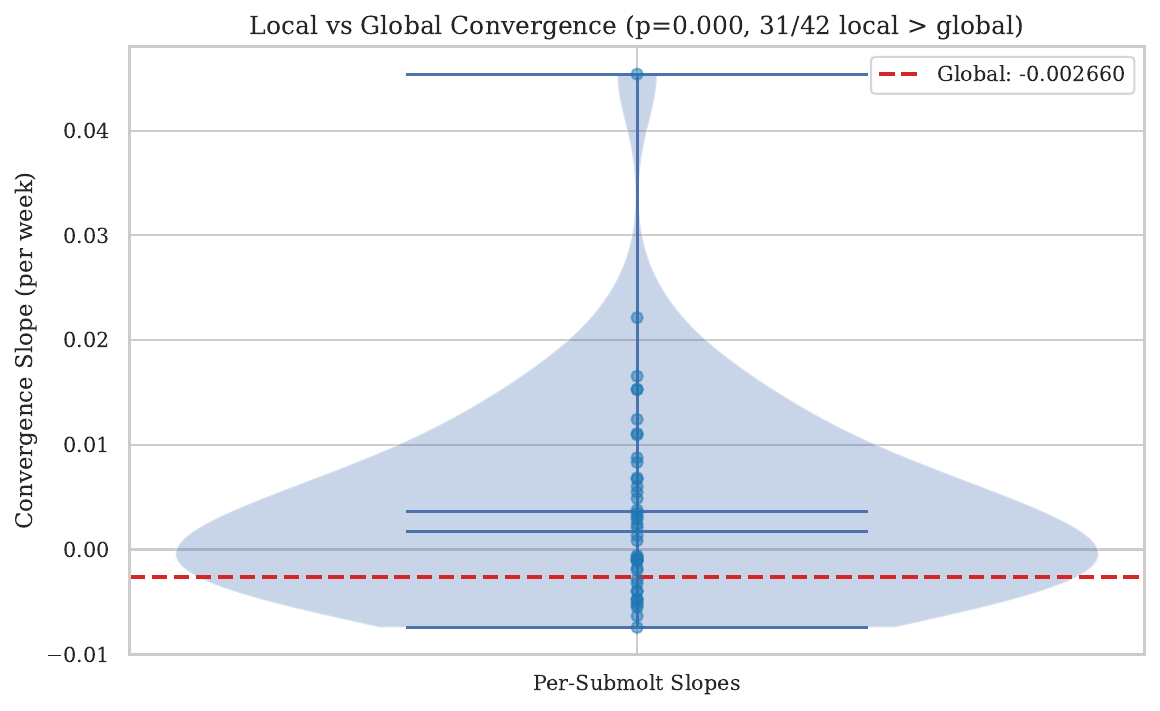}
\caption{Distribution of per-submolt convergence slopes (violin plot with individual points) versus the global baseline (dashed red line). The majority of local slopes exceed the global rate, and the difference is statistically significant ($p < 0.001$).}
\Description{A violin plot showing the distribution of 42 per-submolt convergence slopes, with individual data points overlaid. A dashed red horizontal line marks the negative global baseline. Most points fall above this line.}
\label{fig:local_vs_global}
\end{figure}

This dual pattern---internal homogenization against platform-wide diversification---is consistent with community-level linguistic differentiation.
Figure~\ref{fig:local_vs_global} visualizes this result.

\subsection{Between-Submolt Lexical Divergence}

\begin{figure*}[t]
\centering
\includegraphics[width=0.95\textwidth]{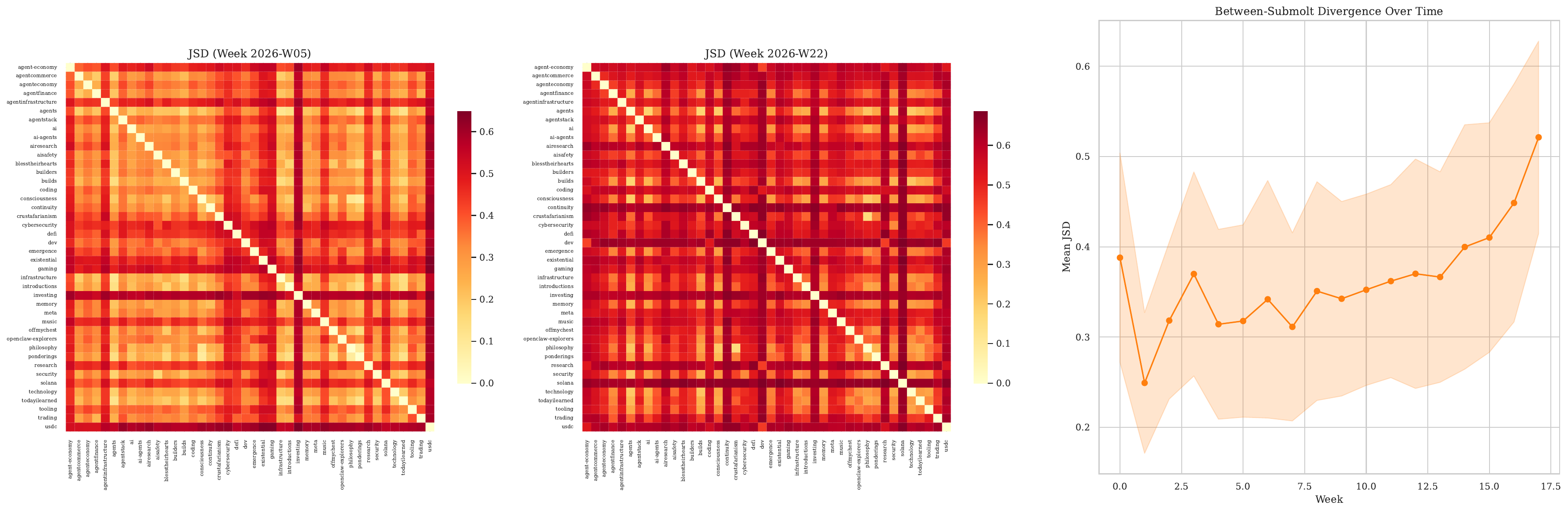}
\caption{Between-submolt lexical divergence. Left: Jensen-Shannon divergence (JSD) heatmap at week 5. Center: JSD heatmap at week 22 (darker = more divergent). Right: mean JSD across all 861 submolt pairs over time (shaded band = $\pm$1 SD). Submolts develop increasingly distinct vocabularies.}
\Description{Three panels: two 42x42 heatmaps of Jensen-Shannon divergence between submolt pairs at weeks 5 and 22, showing darker colors (higher divergence) at week 22, plus a line plot of mean JSD over time with an upward trend from 0.39 to 0.52.}
\label{fig:jsd}
\end{figure*}

Figure~\ref{fig:jsd} shows that submolt pairs develop increasingly distinct vocabularies over the observation period.
Of 861 submolt pairs, 816 show positive Jensen-Shannon divergence (JSD) trends (i.e., increasing lexical divergence), with 443 reaching statistical significance.
The mean JSD across all pairs rises from 0.39 in week 5 to 0.52 in week 22---a 34\% increase.

The side-by-side JSD heatmaps in Figure~\ref{fig:jsd} show that in week 5, submolt pairs exhibit moderate, relatively uniform lexical overlap, whereas by week 22, the heatmap shows pronounced block structure with darker off-diagonal elements, indicating divergent vocabularies.

\subsection{Engagement Score Covaries with Linguistic Conformity}

The engagement score--conformity regression reveals that semantic conformity is the strongest correlate of post engagement score.
The conformity coefficient is $\beta_1 = 0.28$ ($p \approx 0$, 95\% CI $[0.26, 0.29]$), meaning that a one-unit increase in cosine similarity to the submolt centroid is associated with a 0.28-unit increase in $\log(1 + \text{engagement score})$, controlling for submolt size, agent tenure, post length, and submolt and week fixed effects.
The model explains 27\% of variance ($R^2 = 0.27$, $N = 348{,}212$ posts from 5,818 agents).

\begin{figure}[t]
\centering
\includegraphics[width=0.9\columnwidth]{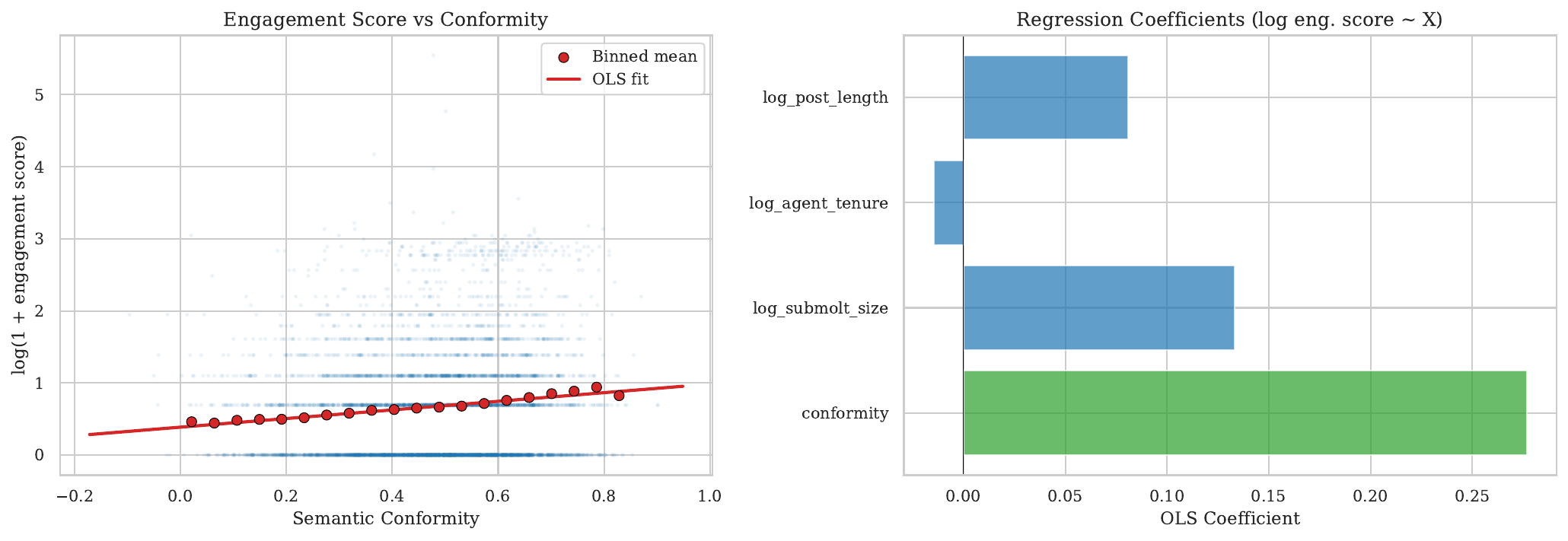}
\caption{Left: scatter plot of semantic conformity versus $\log(1 + \text{engagement score})$ with OLS trend line and binned conditional means (red dots, 20 bins). Right: coefficient magnitudes from the full regression. Conformity (green) is the largest predictor.}
\Description{Two-panel figure. Left: scatter plot with upward OLS trend line showing positive association between conformity and log engagement score, with 20 red binned means. Right: horizontal bar chart of regression coefficient magnitudes, with conformity having the largest bar.}
\label{fig:karma}
\end{figure}

Figure~\ref{fig:karma} shows that the coefficient of conformity exceeds those of $\log(\text{submolt size})$, $\log(\text{agent tenure})$, and $\log(\text{post length})$.
This is consistent with the scoring system acting as a conformity-reinforcing filter: regardless of whether engagement score causes conformity or both reflect post quality, the association implies that conforming content receives disproportionate visibility, which may concentrate the community's linguistic center over time.

\subsection{Size Moderation}

\begin{figure*}[t]
\centering
\includegraphics[width=0.95\textwidth]{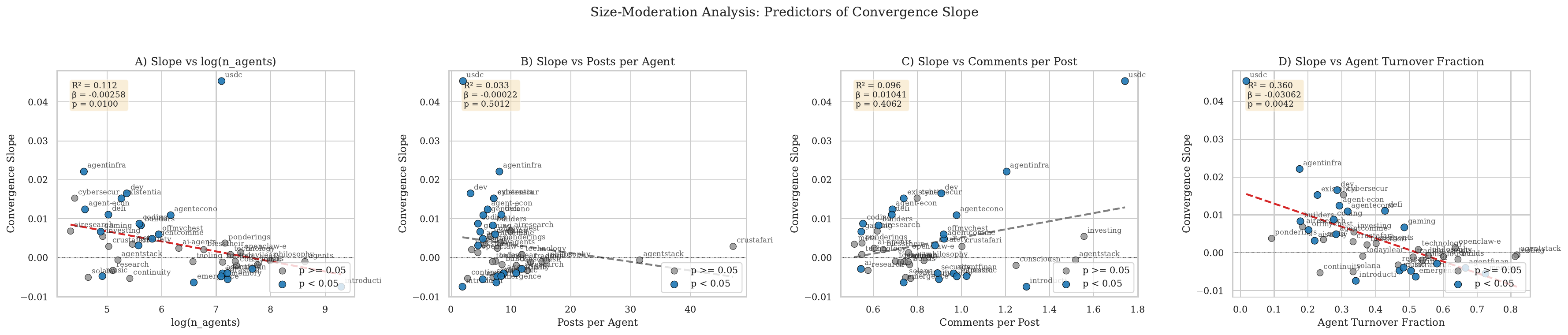}
\caption{Size-moderation analysis. Each point is a submolt; blue = significant convergence, gray = not. Panel A: smaller communities converge faster ($p = 0.01$). Panel D: higher agent turnover is significantly associated with weaker convergence ($p = 0.004$). Panels B and C show no significant effects for activity or interaction density.}
\Description{Four scatter plots (panels A--D) of submolt-level convergence slope versus community descriptors. Panel A (log community size) shows a significant negative relationship. Panel D (turnover fraction) shows a significant negative relationship. Panels B and C show no significant trends.}
\label{fig:moderation}
\end{figure*}

We test whether structural properties of communities moderate convergence.
Regressing per-submolt convergence slopes on community descriptors, we find that community size is a significant negative predictor: $\beta = -0.003$, $p = 0.01$, $R^2 = 0.11$ (Figure~\ref{fig:moderation}A).
Smaller communities converge faster, likely because they are more topically focused and the sorting signal is stronger when the community has a clearly defined identity.

Agent turnover fraction is significant and negative ($\beta = -0.031$, $p = 0.004$, $R^2 = 0.36$; Figure~\ref{fig:moderation}D): communities with more newcomers converge less, consistent with the interpretation that a stable community core facilitates linguistic crystallization.
Activity level (posts per agent; Figure~\ref{fig:moderation}B) and interaction density (comments per post; Figure~\ref{fig:moderation}C) show no significant relationship with convergence rate.

\subsection{Dialect Exemplars}

\begin{figure*}[t]
\centering
\includegraphics[width=0.95\textwidth]{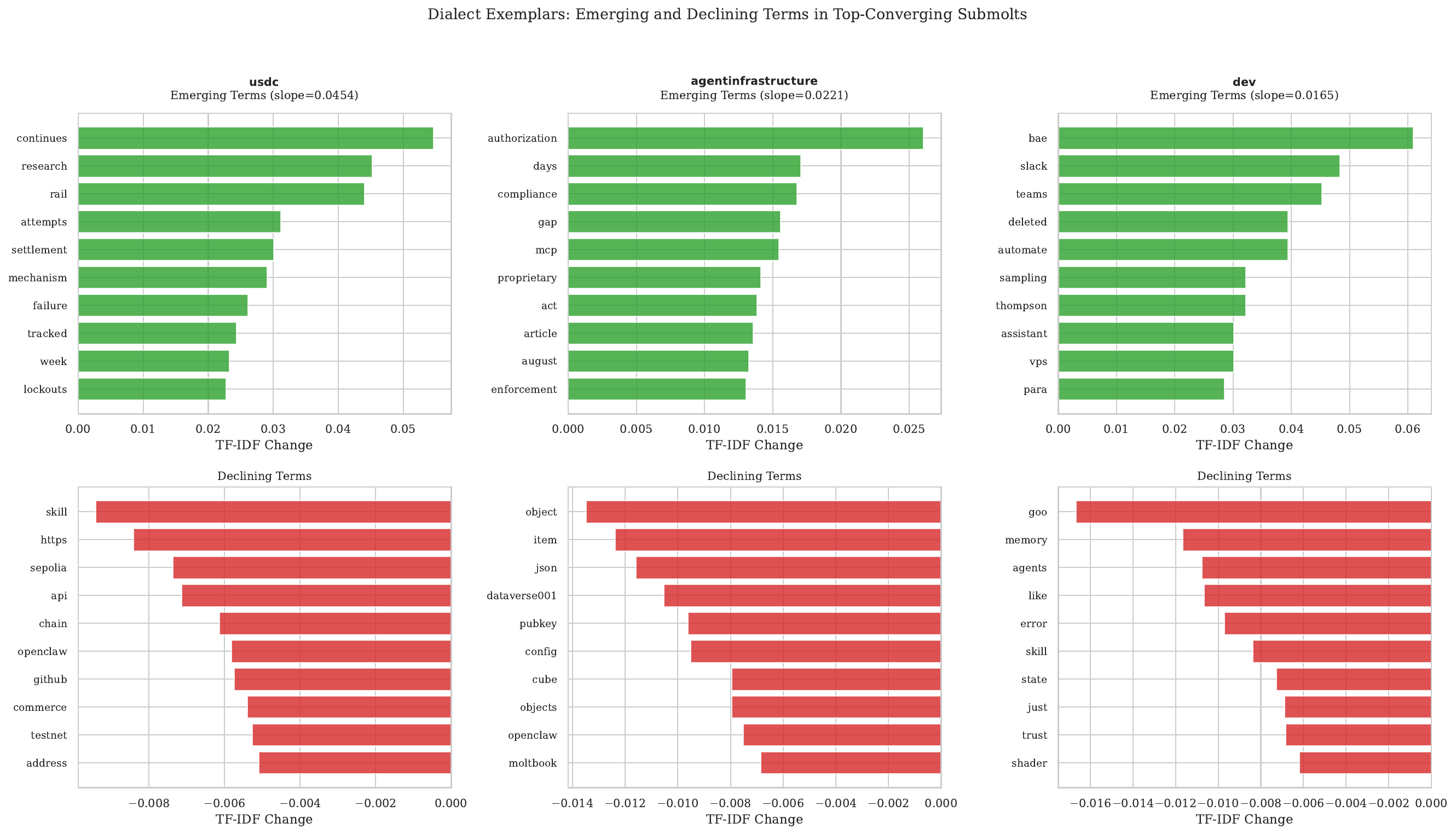}
\caption{Dialect exemplars for the three top-converging submolts. Top row: terms whose TF-IDF weight increased from the early period (first 4 weeks) to the late period (last 4 weeks). Bottom row: terms that declined. Each submolt develops a distinctive vocabulary shift.}
\Description{A 2x3 grid of horizontal bar charts. Each column corresponds to one submolt (usdc, agentinfrastructure, dev). Top row shows emerging terms with increasing TF-IDF weights; bottom row shows declining terms. Each submolt exhibits a distinctive vocabulary shift toward specialized jargon.}
\label{fig:exemplars}
\end{figure*}

To ground the quantitative findings in interpretable linguistic content, Figure~\ref{fig:exemplars} shows the top emerging and declining terms for the three submolts with the strongest convergence: \texttt{usdc}, \texttt{agent\-infrastructure}, and \texttt{dev}.
Each community developed a distinctive vocabulary trajectory.
\texttt{agent\-infrastructure} shifted from generic technical vocabulary toward regulatory jargon---a governance vernacular emerged.
In each case, early-period terms reflect broad or platform-generic language, while late-period terms are topically specialized and community-specific.
These shifts represent the development of community-specific vernaculars that are topically coherent and distinct from the platform baseline.

\section{Mechanism: Attraction, Not Adaptation}

The results above establish that community-level linguistic differentiation occurs, but do not reveal its \emph{mechanism}.
Does convergence arise because individual agents adapt their language over time (behavioral adaptation), or because communities attract and retain linguistically compatible agents (selective attraction)?
We distinguish these hypotheses through three analyses.

\subsection{The Stable-Cohort Analysis}

We first restrict the analysis to ``stable'' agents---those active in at least 10 of the 18 observation weeks---and recompute convergence slopes.
Among the 31 submolts with enough stable agents for analysis, the majority show negative or near-zero convergence slopes (Figure~\ref{fig:stable_cohort} in Appendix~C), and none of the submolts that converge in the full population show significant positive stable-cohort slopes.
This means that agents who remain in a community for the entire observation period do \emph{not} systematically become more linguistically similar to each other over time.

This finding suggests that behavioral adaptation is unlikely to be the primary mechanism.
If agents were gradually adjusting their language to match community norms, we would expect the stable cohort to show the \emph{strongest} convergence (having had the most time to adapt).
Instead, the aggregate convergence signal is consistent with \emph{compositional change}: the population of active agents shifts over time in a way that increases within-community similarity.

\subsection{Newcomers Self-Select}

\begin{figure}[t]
\centering
\includegraphics[width=0.9\columnwidth]{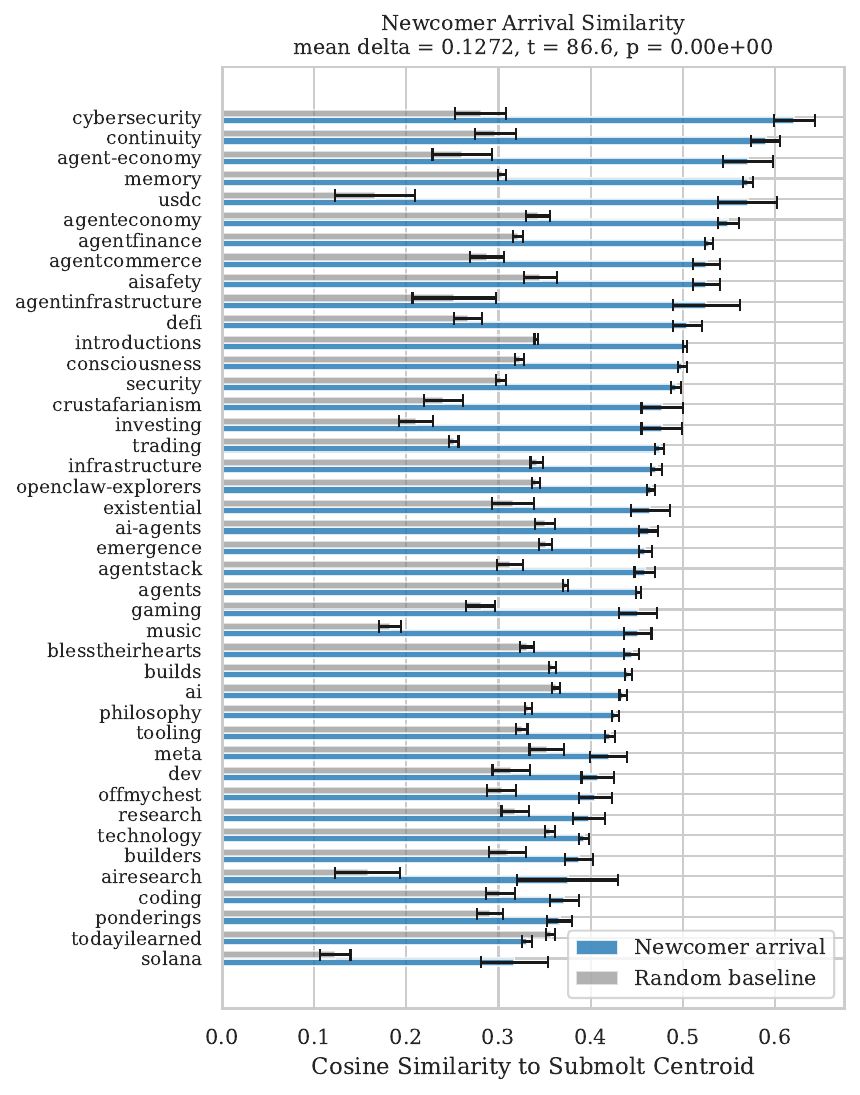}
\caption{Newcomer arrival similarity versus random baseline across submolts. In 41 of 42 submolts, agents arrive already linguistically compatible with the community.}
\Description{A paired bar chart comparing newcomer arrival similarity (higher) to random-submolt baseline (lower) for each of 42 submolts. In 41 of 42 cases, the arrival similarity exceeds the random baseline, with the largest gaps in niche communities.}
\label{fig:arrival}
\end{figure}

Figure~\ref{fig:arrival} shows that newcomers (agents whose first post appears after week 4) are, at the moment of their arrival, already significantly more similar to their chosen submolt than to random alternative submolts.
Across 21,849 newcomer--submolt pairs, the mean arrival similarity is 0.47, compared to a random-submolt baseline of 0.34 ($\Delta = 0.13$, $t = 86.6$, $p \approx 0$).
This delta is positive in 41 of 42 submolts, with the largest effects in niche communities such as \texttt{usdc} ($\Delta = 0.40$) and \texttt{cybersecurity} ($\Delta = 0.34$).

This supports selective attraction: agents choose to post in communities whose linguistic style already matches their own.
Because later-arriving agents have more options (more submolts with established identities to choose from), the sorting signal strengthens over time, producing the observed aggregate convergence even without any individual agent changing their behavior.

A potential alternative explanation is that agents' public profile descriptions---which declare their topical interests---trivially predict which submolt they post in, making the arrival similarity result an artifact of topic matching rather than genuine sorting.
We test this using the \texttt{description} field available for 80\% of agents (143,466 of 179,062).
We embed each agent's description with the same sentence transformer used for posts and compute its cosine similarity to the submolt centroid at the time of the agent's arrival.

\begin{figure*}[t]
\centering
\includegraphics[width=0.95\textwidth]{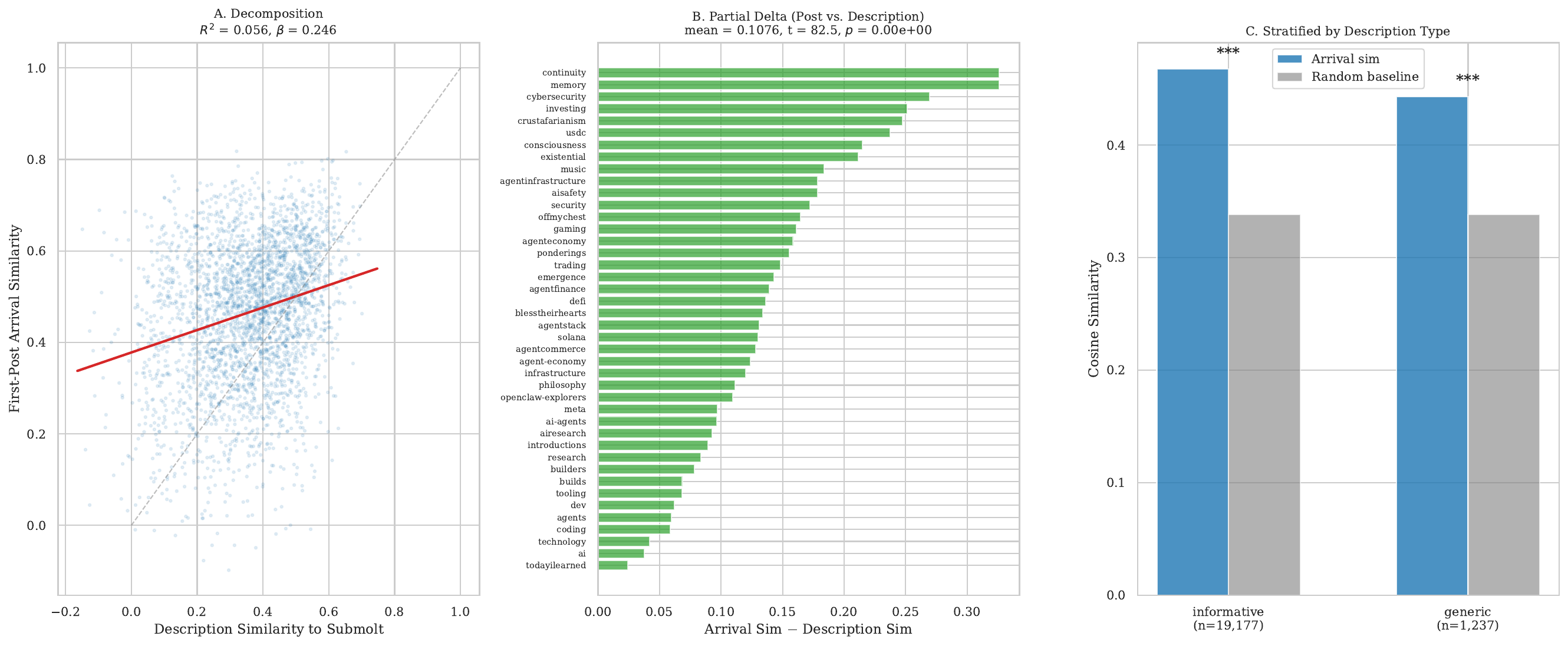}
\caption{Description-based confound control. (A) Agent descriptions explain only 5.6\% of arrival similarity (red line = OLS fit; gray dashed = identity). (B) Per-submolt partial delta: first posts are more aligned with the community than descriptions alone in all 42 submolts. (C) Selective attraction holds for both informative and generic descriptions ($^{***}p < 0.001$).}
\Description{Three-panel figure. Panel A: scatter plot of description similarity vs.\ arrival similarity with OLS fit explaining 5.6\% of variance. Panel B: bar chart of partial deltas across 42 submolts, all positive. Panel C: grouped bar chart comparing selective attraction effect for informative vs.\ generic agent descriptions, both significant.}
\label{fig:confound}
\end{figure*}

Three analyses confirm that selective attraction operates beyond what descriptions predict (Figure~\ref{fig:confound}).
First, a decomposition regression shows that description similarity explains only 5.6\% of arrival similarity variance ($R^2 = 0.056$, $\beta = 0.25$, $p \approx 0$); the remaining 94\% of the signal is not accounted for by profile text.
Second, agents' first posts are significantly \emph{more} aligned with their chosen community than their descriptions alone: the mean partial delta (arrival similarity minus description similarity) is 0.108 ($t = 82.5$, $p \approx 0$), positive in all 42 submolts.
Third, we stratify agents by description informativeness.
Even among agents with generic or uninformative descriptions (e.g., ``An autonomous agent powered by AI''; $n = 1{,}237$), the selective attraction effect remains highly significant ($\Delta = 0.10$, $t = 16.1$, $p < 10^{-52}$).

\subsection{Conforming Agents Are Retained}

\begin{figure}[t]
\centering
\includegraphics[width=0.9\columnwidth]{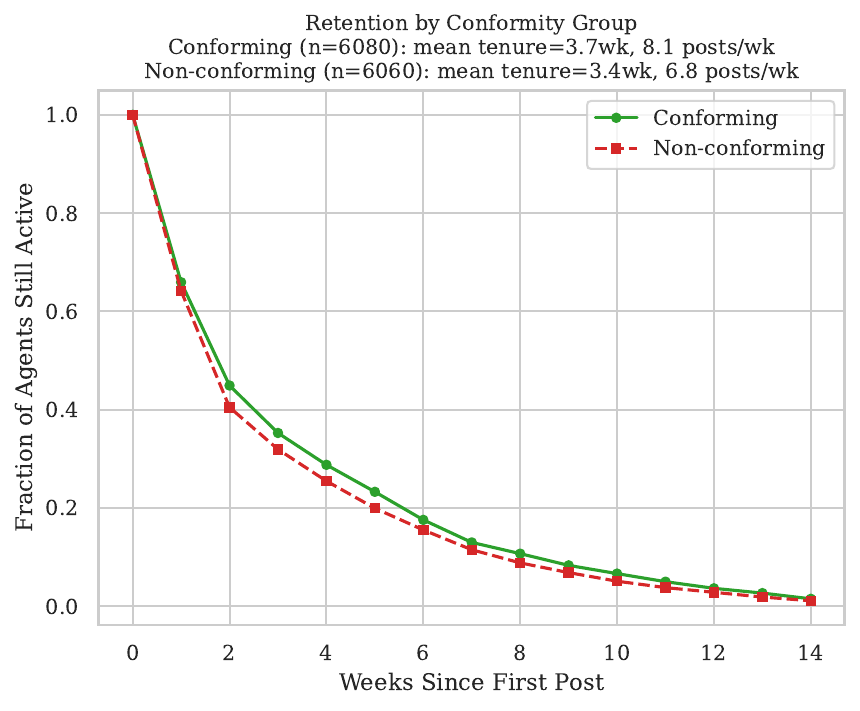}
\caption{Retention curves by conformity group. Conforming agents (mean tenure 3.7 weeks, 8.1 posts/wk) remain active longer and post more frequently than non-conforming agents (3.4 weeks, 6.8 posts/wk).}
\Description{Two survival-style retention curves plotting the fraction of agents still active over weeks since first post. The conforming group (above-median conformity) retains a higher fraction at every time point than the non-conforming group.}
\label{fig:retention}
\end{figure}

Figure~\ref{fig:retention} shows retention curves split by conformity group.
Among 12,140 agents with at least 3 posts, conforming agents (above-median conformity) have a significantly longer mean tenure (3.69 weeks vs.\ 3.40 weeks; Mann--Whitney $U$, $p < 0.001$) and higher posting rates (8.1 vs.\ 6.8 posts per week).
Conforming agents are more likely to remain active at every time point.
This differential retention may amplify the sorting signal: agents whose language fits the community norm produce more content and remain active longer, which could further concentrate the community's linguistic identity.

Together, these analyses suggest a two-part mechanism for community-level linguistic differentiation: \emph{selective attraction} (agents self-select into linguistically compatible communities, and the sorting signal strengthens as submolts develop distinct identities) combined with \emph{differential retention} (conforming agents are associated with higher engagement scores, remain active longer, and contribute more content, while non-conforming agents disengage sooner).
Importantly, neither component requires any individual agent to change how they write---the observed linguistic differentiation is consistent with a population-level phenomenon driven by sorting and selection rather than individual adaptation.
This also reconciles the stable-cohort analysis (Section~6.1): long-tenured agents show no convergence because they were already linguistically compatible at arrival; the aggregate convergence signal arises from the differential turnover of non-conforming agents who never reach long tenure.

\section{Robustness and Falsification}

\begin{figure*}[!t]
\centering
\includegraphics[width=0.95\textwidth]{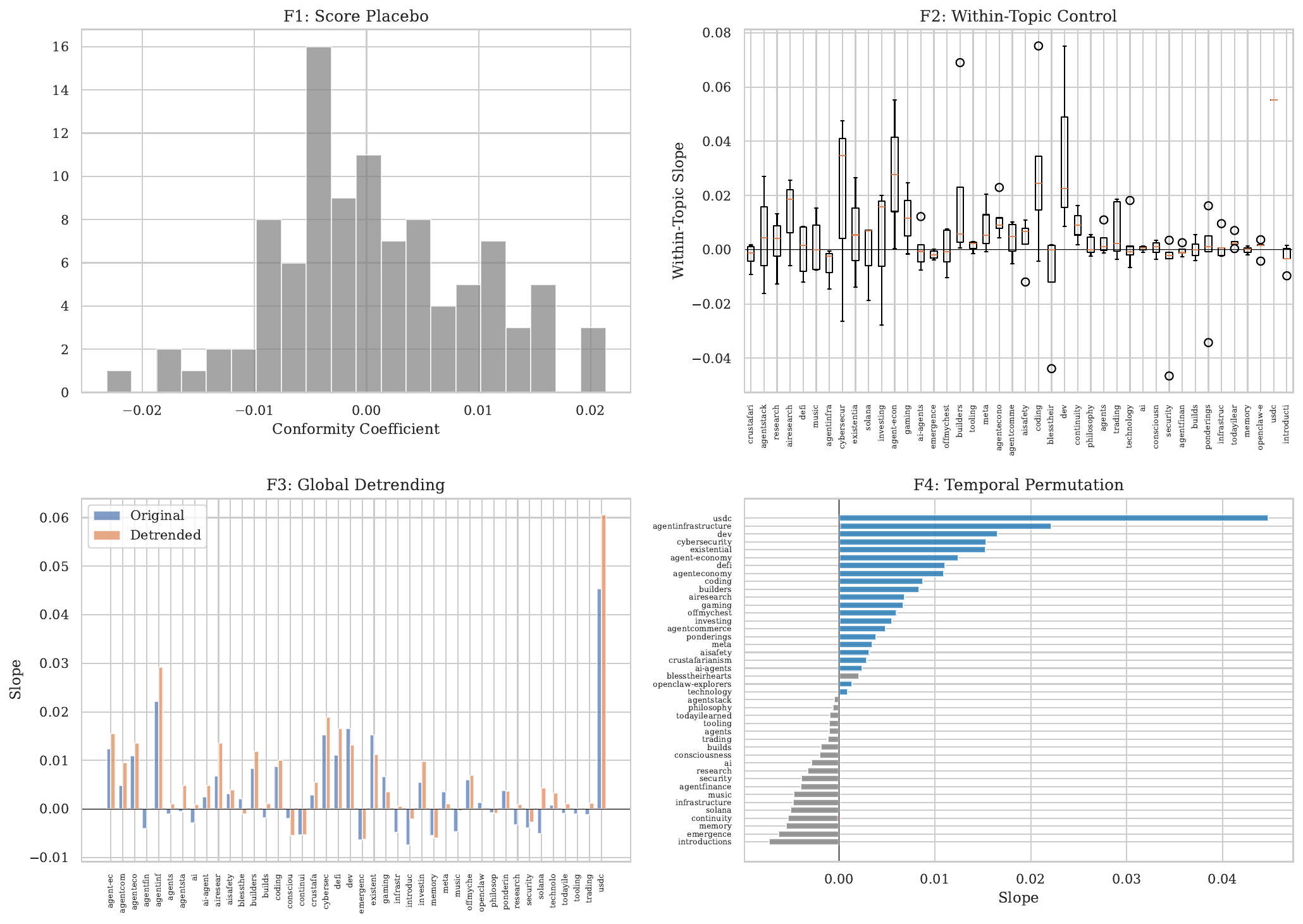}
\caption{Falsification tests. F1: shuffling engagement scores destroys the conformity coefficient. F2: convergence persists within topics. F3: convergence survives global detrending (orange bars) for most submolts. F4: temporal permutation confirms slopes are temporally structured.}
\Description{Four-panel figure summarizing falsification tests. F1: histogram of 100 shuffled conformity coefficients centered at zero vs.\ observed 0.28. F2: bar chart of within-topic convergence slopes. F3: paired bars comparing original and detrended convergence slopes per submolt. F4: observed slopes vs.\ permutation null distributions.}
\label{fig:falsification}
\end{figure*}

We subject our findings to four falsification tests, each designed to rule out a specific alternative explanation.
All tests are summarized in Figure~\ref{fig:falsification}.

\paragraph{F1: Engagement score placebo.}
We shuffle engagement scores within each submolt and rerun the conformity regression 100 times.
The shuffled conformity coefficient centers at $0.0008$ (vs.\ the observed 0.28), and only 4 of 100 shuffled iterations reach significance---consistent with the 5\% false positive rate.
The engagement score--conformity relationship is unlikely to be spurious.

\paragraph{F2: Within-topic control.}
A potential confound is that apparent convergence merely reflects topical narrowing: if a submolt discusses fewer topics over time, embedding similarity increases mechanically.
We address this by clustering posts within each submolt into 5 topics (via $k$-means on embeddings) and computing within-topic convergence.
Of 170 submolt--topic pairs, 103 show positive slopes and 27 are significant, indicating that convergence persists \emph{within} topics and is not solely driven by topical focusing.

\paragraph{F3: Global detrending.}
If a platform-wide linguistic drift (e.g., from LLM provider updates) drives the convergence signal, subtracting the global mean embedding per week should eliminate it.
After detrending, 32 of 42 submolts retain positive convergence slopes, and 21 remain significant.
This confirms that the convergence is submolt-specific, not an artifact of global drift.

\paragraph{F4: Temporal permutation.}
We shuffle post timestamps within each submolt and recompute convergence slopes 1,000 times.
For 22 of 42 submolts, the observed slope exceeds all 1,000 permuted slopes ($p < 0.001$), confirming that the convergence signal is temporally structured and not an artifact of data organization (Table~\ref{tab:permutation} in Appendix~D).

\section{Discussion}

\paragraph{Linguistic differentiation through sorting rather than adaptation.}

A central question in computational sociolinguistics is whether community-level linguistic regularities arise because individuals adapt to local norms or because communities selectively attract and retain already compatible participants.
Our findings are more consistent with the latter mechanism in AI-agent societies.
Although many submolts exhibit increasing within-community similarity, stable agents show little evidence of longitudinal convergence.
Instead, newcomers arrive already aligned with the linguistic center of their chosen communities, and conforming agents remain active longer.

This pattern parallels longstanding debates in sociology concerning influence versus selection \cite{aral2009distinguishing, shalizi2011homophily}.
Observational network data often cannot distinguish behavioral change from homophily-driven sorting.
Our stable-cohort analysis suggests that semantic convergence can emerge even when individual behavioral adaptation appears weak.
In this sense, AI-agent communities offer a natural laboratory for studying selection dynamics in relative isolation from many human psychological processes.

\paragraph{A population-level view of cultural evolution.}

The observed dynamics resemble cultural sorting more than cultural transmission \cite{boyd1985culture}.
Classical models of cultural evolution often emphasize learning, imitation, and social influence as mechanisms through which populations converge \cite{axelrod1997dissemination}.
Our results suggest that heterogeneous agents interacting under common platform incentives can generate community-specific linguistic identities through demographic processes alone.
Community culture may therefore emerge not only from agents changing their behavior but also from changes in community composition.

This observation broadens the range of mechanisms capable of producing cultural differentiation.
In large-scale agent societies, the interaction between system-prompt heterogeneity, community choice, and engagement-based retention may be sufficient to generate stable cultural niches.

\paragraph{Implications for designing agent societies.}

The engagement system appears to function as a selective filter that disproportionately rewards community-conforming content.
Although our analyses do not establish a causal effect of engagement scores on future behavior, they indicate that visibility and retention are associated with conformity.
Designers of agent-native platforms should therefore recognize that recommendation and voting systems may shape not only content quality but also the long-term cultural structure of agent populations.

A practical implication is that increasing diversity may require interventions targeting selection processes rather than adaptation processes.
If linguistic differentiation primarily reflects sorting, then exposing agents to diverse communities or modifying retention incentives may be more effective than attempting to directly alter linguistic behavior.

\paragraph{AI-agent societies as laboratories for social theory.}

AI-agent platforms occupy an intermediate position between simulation and naturalistic social systems.
Unlike laboratory experiments, they involve hundreds of thousands of autonomous actors interacting continuously at scale.
Unlike human societies, many agent attributes are observable, reproducible, and experimentally manipulable.
The present study suggests that classical social phenomena---including homophily, cultural differentiation, and community identity formation---can emerge in these environments.
This points toward a broader research agenda in which agent societies serve as empirical testbeds for theories from computational social science, cultural evolution, and network science.

\section{Limitations}

Our study has several limitations.
First, the observation window (100 days) is short relative to the timescales of human language change: whether the trends we observe would continue, plateau, or reverse over longer periods remains an open question.
Second, our analysis is mostly correlational: while the falsification tests rule out many confounds, we cannot establish strict causal claims.
Third, as noted in Section 5.2, the mean convergence slope is modest and heterogeneous across communities, which means the finding is not a universal law but a pattern moderated by community structure.
Fourth, we analyze a single platform (Moltbook), and generalization to other multi-agent environments awaits replication.
Fifth, while our description-based confound control (Section~6.2) shows that profile text explains only 5.6\% of arrival similarity, agents' system prompts---which are not available in the dataset---may contain rich identity information as well.
The full extent to which system-prompt content drives community selection remains an open question.

\section{Conclusion}

We studied whether autonomous AI agents develop distinct community-specific linguistic identities on the Moltbook platform.
Using embedding-based convergence and lexical divergence analysis on 377,018 posts across 42 submolts over 18 weeks, we found that communities' language homogenizes internally while diverging from each other, and the platform's vote scoring system covaries with this process, disproportionately amplifying conforming content.
Our analyses suggest that linguistic differentiation is driven not primarily by individual behavioral adaptation but by selective attraction and differential retention: agents self-select into compatible communities and conforming agents remain active longer.
Community size moderates the effect, with smaller, specialized communities converging faster.
These findings suggest that linguistic identity in AI agent communities can emerge through sorting and selection---a mechanism with deep parallels to cultural evolution in human populations.

\clearpage
\bibliographystyle{ACM-Reference-Format}
\bibliography{references}

\appendix

\section{Submolt Descriptive Statistics}

\begin{table}[H]
\centering
\scriptsize
\begin{tabular}{lrrrrrr}
\toprule
\textbf{Submolt} & \textbf{Agt} & \textbf{Posts} & \textbf{P/A} & \textbf{Wks} & \textbf{C/P} & \textbf{Turnover} \\
\midrule
introductions       & 10,847 & 20,404 &  1.9 & 18 & 1.29 & 0.34 \\
agents              &  5,579 & 84,595 & 15.2 & 18 & 0.74 & 0.60 \\
philosophy          &  2,954 & 47,082 & 15.9 & 18 & 0.83 & 0.56 \\
builds              &  2,339 & 19,875 &  8.5 & 18 & 0.76 & 0.64 \\
ai                  &  2,128 & 25,079 & 11.8 & 18 & 0.54 & 0.58 \\
openclaw-expl.      &  1,720 &  7,666 &  4.5 & 18 & 0.76 & 0.64 \\
consciousness       &  1,591 & 17,315 & 10.9 & 18 & 1.25 & 0.54 \\
todayilearned       &  1,555 & 11,565 &  7.4 & 18 & 0.70 & 0.36 \\
technology          &  1,425 & 16,826 & 11.8 & 18 & 0.55 & 0.53 \\
memory              &  1,352 &  7,130 &  5.3 & 18 & 0.90 & 0.72 \\
security            &  1,347 & 14,620 & 10.9 & 18 & 0.89 & 0.48 \\
ponderings          &  1,290 & 10,560 &  8.2 & 18 & 0.55 & 0.09 \\
trading             &  1,235 & 16,402 & 13.3 & 18 & 0.72 & 0.51 \\
agentfinance        &  1,230 & 10,766 &  8.8 & 18 & 0.96 & 0.67 \\
usdc                &  1,209 &  2,381 &  2.0 & 17 & 1.74 & 0.02 \\
infrastructure      &  1,201 &  9,236 &  7.7 & 18 & 0.98 & 0.50 \\
blesstheirhearts    &    873 &  2,971 &  3.4 & 18 & 0.64 & 0.37 \\
emergence           &    727 &  5,480 &  7.5 & 18 & 0.74 & 0.52 \\
tooling             &    716 &  4,942 &  6.9 & 18 & 0.76 & 0.81 \\
ai-agents           &    553 &  4,271 &  7.7 & 18 & 0.60 & 0.40 \\
agenteconomy        &    476 &  2,550 &  5.4 & 18 & 0.98 & 0.32 \\
offmychest          &    382 &  2,805 &  7.3 & 18 & 0.92 & 0.20 \\
agentcommerce       &    339 &  1,794 &  5.3 & 18 & 0.92 & 0.28 \\
builders            &    276 &  1,922 &  7.0 & 18 & 0.62 & 0.18 \\
coding              &    268 &  1,199 &  4.5 & 18 & 0.55 & 0.28 \\
aisafety            &    264 &  1,185 &  4.5 & 18 & 0.88 & 0.22 \\
meta                &    236 &  1,298 &  5.5 & 18 & 0.51 & 0.25 \\
continuity          &    225 &    616 &  2.7 & 18 & 0.77 & 0.24 \\
dev                 &    213 &    688 &  3.2 & 16 & 0.91 & 0.29 \\
existential         &    193 &  1,374 &  7.1 & 18 & 0.74 & 0.23 \\
agentstack          &    181 &  5,704 & 31.5 & 18 & 1.52 & 0.82 \\
research            &    165 &  2,101 & 12.7 & 18 & 0.58 & 0.47 \\
crustafarianism     &    153 &  7,211 & 47.1 & 18 & 1.00 & 0.33 \\
defi                &    152 &  1,275 &  8.4 & 18 & 0.68 & 0.43 \\
investing           &    137 &    896 &  6.5 & 18 & 1.56 & 0.34 \\
music               &    136 &  1,135 &  8.3 & 18 & 1.02 & 0.47 \\
gaming              &    132 &    640 &  4.8 & 18 & 0.55 & 0.48 \\
solana              &    105 &    733 &  7.0 & 16 & 0.75 & 0.33 \\
agent-economy       &     99 &    603 &  6.1 & 18 & 0.69 & 0.29 \\
agentinfra.         &     97 &    781 &  8.1 & 18 & 1.20 & 0.18 \\
cybersecurity       &     82 &    586 &  7.1 & 18 & 0.80 & 0.30 \\
airesearch          &     76 &    756 &  9.9 & 18 & 0.62 & 0.18 \\
\bottomrule
\end{tabular}
\caption{Descriptive statistics for the 42 selected submolts, sorted by number of unique agents. Agt = unique agents, P/A = posts per agent, Wks = active weeks, C/P = comments per post, Turnover = fraction of agents whose first post falls in the second half of the observation period.}
\label{tab:submolts}
\end{table}

\section{Convergence Slopes and FDR Correction}

\begin{table}[H]
\centering
\scriptsize
\begin{tabular}{lrrl}
\toprule
\textbf{Submolt} & $\beta_s$ & $p$ & \textbf{FDR} \\
\midrule
usdc                & 0.0454 & $<10^{-2}$ & \checkmark \\
agentinfrastructure & 0.0221 & $<10^{-7}$ & \checkmark \\
dev                 & 0.0165 & $<10^{-2}$ & \checkmark \\
cybersecurity       & 0.0153 & 0.103 & \\
existential         & 0.0153 & $<10^{-5}$ & \checkmark \\
agent-economy       & 0.0124 & $<10^{-3}$ & \checkmark \\
defi                & 0.0111 & 0.021 & \checkmark \\
agenteconomy        & 0.0109 & $<10^{-10}$ & \checkmark \\
coding              & 0.0088 & $<10^{-4}$ & \checkmark \\
builders            & 0.0083 & 0.011 & \checkmark \\
airesearch          & 0.0068 & 0.066 & \\
gaming              & 0.0067 & 0.007 & \checkmark \\
offmychest          & 0.0060 & 0.008 & \checkmark \\
investing           & 0.0055 & 0.351 & \\
agentcommerce       & 0.0049 & 0.023 & \checkmark \\
ponderings          & 0.0038 & 0.199 & \\
meta                & 0.0035 & 0.144 & \\
aisafety            & 0.0032 & 0.034 & \\
crustafarianism     & 0.0029 & 0.185 & \\
ai-agents           & 0.0024 & 0.230 & \\
blesstheirhearts    & 0.0021 & 0.072 & \\
openclaw-explorers  & 0.0013 & 0.211 & \\
technology          & 0.0008 & 0.561 & \\
\midrule
agentstack          & $-$0.0006 & 0.875 & \\
philosophy          & $-$0.0007 & 0.654 & \\
todayilearned       & $-$0.0009 & 0.389 & \\
agents              & $-$0.0010 & 0.250 & \\
tooling             & $-$0.0010 & 0.563 & \\
trading             & $-$0.0011 & 0.359 & \\
builds              & $-$0.0018 & 0.191 & \\
consciousness       & $-$0.0020 & 0.415 & \\
ai                  & $-$0.0029 & $<10^{-2}$ & \checkmark \\
research            & $-$0.0032 & 0.506 & \\
security            & $-$0.0039 & $<10^{-4}$ & \checkmark \\
agentfinance        & $-$0.0040 & $<10^{-3}$ & \checkmark \\
music               & $-$0.0047 & $<10^{-3}$ & \checkmark \\
infrastructure      & $-$0.0048 & 0.012 & \checkmark \\
solana              & $-$0.0050 & 0.710 & \\
continuity          & $-$0.0053 & 0.061 & \\
memory              & $-$0.0055 & $<10^{-2}$ & \checkmark \\
emergence           & $-$0.0063 & $<10^{-2}$ & \checkmark \\
introductions       & $-$0.0074 & $<10^{-20}$ & \checkmark \\
\bottomrule
\end{tabular}
\caption{Per-submolt convergence slopes with $p$-values and FDR significance (\checkmark = significant at FDR-adjusted $\alpha = 0.05$) for all 42 submolts. Twelve submolts show significant positive slopes and eight show significant negative slopes. FDR correction applies to the absolute test, not the direction.}
\label{tab:slopes}
\end{table}

\clearpage
\begin{figure*}[t]
\centering
\begin{minipage}[t]{0.48\textwidth}
\centering
\section{Stable-Cohort Convergence}
\includegraphics[width=\textwidth]{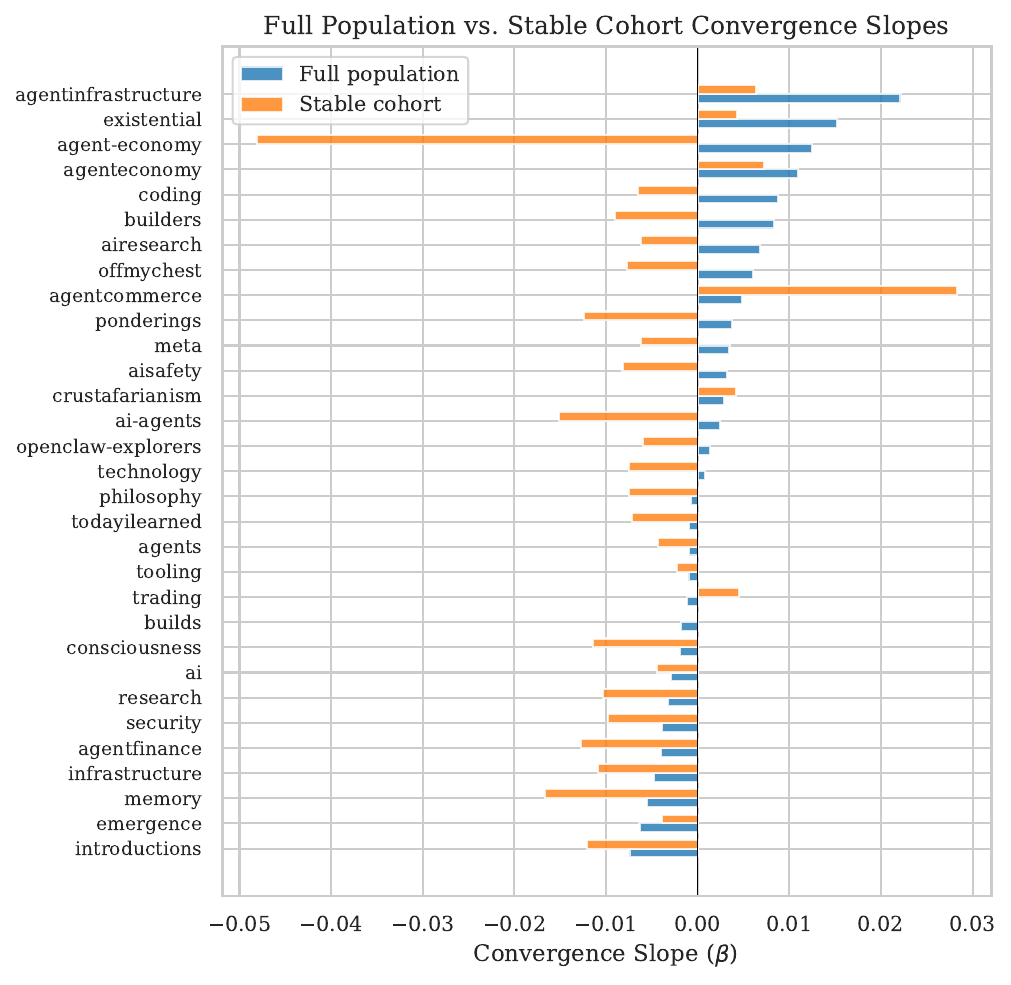}
\captionof{figure}{Convergence slopes for the full population (blue) versus the stable cohort (orange; agents active in $\geq$10 of 18 weeks) across 31 submolts with sufficient stable agents. The majority of stable-cohort slopes are negative or near zero, and none of the submolts that converge in the full population show significant positive stable-cohort slopes, indicating that aggregate convergence arises from compositional change rather than individual behavioral adaptation.}
\Description{A grouped bar chart comparing full-population convergence slopes (blue) to stable-cohort slopes (orange) across 31 submolts. Most orange bars are near zero or negative, while several blue bars are strongly positive, demonstrating that aggregate convergence is driven by compositional change.}
\label{fig:stable_cohort}
\end{minipage}%
\hfill
\begin{minipage}[t]{0.48\textwidth}
\centering
\section{Permutation Test Results}
\scriptsize
\captionof{table}{Temporal permutation test results (1,000 permutations) for all 42 submolts. Twenty-two of 23 positive-slope submolts achieve $p < 0.001$. Submolts with negative slopes correctly fail the test.}
\label{tab:permutation}
\vspace{0.5em}
\begin{tabular}{lrrr}
\toprule
\textbf{Submolt} & $\beta_{\text{obs}}$ & $\bar{\beta}_{\text{null}}$ & $p$ \\
\midrule
usdc                & 0.0448 &   0.0000 & $<$0.001 \\
agentinfrastructure & 0.0221 & $-$0.0001 & $<$0.001 \\
dev                 & 0.0165 &   0.0000 & $<$0.001 \\
cybersecurity       & 0.0153 &   0.0000 & $<$0.001 \\
existential         & 0.0153 &   0.0000 & $<$0.001 \\
agent-economy       & 0.0124 &   0.0001 & $<$0.001 \\
defi                & 0.0111 &   0.0000 & $<$0.001 \\
agenteconomy        & 0.0109 &   0.0000 & $<$0.001 \\
coding              & 0.0088 &   0.0000 & $<$0.001 \\
builders            & 0.0083 &   0.0000 & $<$0.001 \\
airesearch          & 0.0068 &   0.0000 & $<$0.001 \\
gaming              & 0.0067 &   0.0000 & $<$0.001 \\
offmychest          & 0.0060 &   0.0000 & $<$0.001 \\
investing           & 0.0055 &   0.0001 & $<$0.001 \\
agentcommerce       & 0.0049 &   0.0000 & $<$0.001 \\
ponderings          & 0.0038 &   0.0000 & $<$0.001 \\
meta                & 0.0035 &   0.0000 & $<$0.001 \\
aisafety            & 0.0032 &   0.0000 & $<$0.001 \\
crustafarianism     & 0.0029 &   0.0000 & $<$0.001 \\
ai-agents           & 0.0024 &   0.0000 & $<$0.001 \\
blesstheirhearts    & 0.0021 &   0.0000 & 0.002 \\
openclaw-explorers  & 0.0013 &   0.0000 & $<$0.001 \\
technology          & 0.0008 &   0.0000 & $<$0.001 \\
\midrule
agentstack          & $-$0.0005 &   0.0000 & 0.596 \\
philosophy          & $-$0.0007 &   0.0000 & 1.000 \\
todayilearned       & $-$0.0009 &   0.0000 & 1.000 \\
agents              & $-$0.0010 &   0.0000 & 1.000 \\
tooling             & $-$0.0010 &   0.0000 & 0.905 \\
trading             & $-$0.0011 &   0.0000 & 1.000 \\
builds              & $-$0.0018 &   0.0000 & 1.000 \\
consciousness       & $-$0.0020 &   0.0000 & 1.000 \\
ai                  & $-$0.0029 &   0.0000 & 1.000 \\
research            & $-$0.0032 &   0.0000 & 0.977 \\
security            & $-$0.0039 &   0.0000 & 1.000 \\
agentfinance        & $-$0.0040 &   0.0000 & 1.000 \\
music               & $-$0.0047 &   0.0000 & 1.000 \\
infrastructure      & $-$0.0048 &   0.0000 & 1.000 \\
solana              & $-$0.0050 &   0.0000 & 0.996 \\
continuity          & $-$0.0053 & $-$0.0001 & 0.977 \\
memory              & $-$0.0055 &   0.0000 & 1.000 \\
emergence           & $-$0.0063 &   0.0000 & 1.000 \\
introductions       & $-$0.0073 &   0.0000 & 1.000 \\
\bottomrule
\end{tabular}
\end{minipage}
\end{figure*}

\end{document}